\documentclass[aps,reprint,floatfix]{revtex4-2}
\usepackage[utf8]{inputenc}
\usepackage[T1]{fontenc}
\usepackage{enumerate}
\usepackage{enumitem}

\usepackage{amsmath}
\usepackage{amssymb}
\usepackage{amsfonts}
\usepackage{graphicx}
\usepackage{siunitx}
\usepackage{xcolor}
\usepackage{physics}

\usepackage[colorlinks=true,citecolor=blue,urlcolor=red]{hyperref}

\newcommand{\K}{K}
\newcommand{\W}{\mathcal{W}}
\newcommand{\B}{\mathcal{B}}
\newcommand{\x}{\mathbf{x}}
\newcommand{\X}{\mathbf{X}}

\newcommand{\methods}{\emph{Methods}}

\newcommand{\Q}{\mathcal{Q}}
\newcommand{\V}{\mathcal{V}}

\begin{document}

\title{Reliability and operation cost of underdamped memories during cyclic erasures}
\author{Salamb\^{o} Dago}
\author{Sergio Ciliberto}
\author{Ludovic Bellon}
\email{ludovic.bellon@ens-lyon.fr}
\affiliation{Univ Lyon, ENS de Lyon, CNRS, Laboratoire de Physique, F-69342 Lyon, France}
\begin{abstract}
The reliability of fast repeated erasures is studied experimentally and theoretically in a 1-bit underdamped memory. The bit is encoded by the position of a micro-mechanical oscillator whose motion is confined in a double well potential. To contain the energetic cost of fast erasures, we use a resonator with high quality factor $Q$: the erasure work $\W$ is close to Landauer's bound, even at high speed. The drawback is the rise of the system's temperature $T$ due to a weak coupling to the environment. Repeated erasures without letting the memory thermalize between operations result in a continuous warming, potentially leading to a thermal noise overcoming the barrier between the potential wells. In such case, the reset operation can fail to reach the targeted logical state. The reliability is characterized by the success rate $R^{\textrm{s}}_i$ after $i$ successive operations. $\W$, $T$ and $R^{\textrm{s}}_i$ are studied experimentally as a function of the erasure speed. Above a velocity threshold, $T$ soars while $R^{\textrm{s}}_i$ collapses: the reliability of too fast erasures is low. These experimental results are fully justified by two complementary models. We demonstrate that $Q\simeq 10$ is optimal to contain energetic costs and maintain high reliability standards for repeated erasures at any speed.
\end{abstract}

\maketitle

The performance of information storage and processing is not only bounded by technological advances, but also constrained by fundamental physics laws: handling information requires energy~\cite{Landauer-1961,Parrondo_sagawa}. R. Landauer laid the foundations for the connection between information theory and thermodynamics by demonstrating theoretically the lower bound required to erase a one-bit memory: $W_{LB}=k_BT_0 \ln2=\SI{3e-21}{J}$ at room temperature $T_0$~\cite{Landauer-1961}, with $k_B$ Boltzmann's constant. This tiny limit has been experimentally illustrated since, using quasi-static processes in model experiments~\cite{Berut2012, Berut2015, orl12, Bech2014, Gavrilov_EPL_2016, Finite_time_2020, Hong_nano_2016, mar16, Dago-2021}. Paving the way to concrete applications, many researches tackled information processing in a finite time. It was found that when decreasing the duration of operations, an energy overhead proportional to the processing speed appears~\cite{Finite_time_2020, Berut2012,Aurell_2012,sek66,PhysRevE.92.032117} and could explain why nowadays fast devices still consume orders of magnitude more energy than Landauer's bound.

Several strategies have been explored to decrease this extra energy consumption: use intrinsically fast devices~\cite{orl12}, lower the damping mechanism~\cite{Dago-2021}, use an out of equilibrium final state~\cite{Finite_time_2020}. However, in this perspective, only the thermodynamic cost of a \emph{single} erasure in a finite time has been studied, allowing an infinitely long relaxation afterwards: as long as the final state reaches the targeted logical state with high fidelity, no prescription on the equilibration time of the system has been studied. In this article, we aim to go beyond this fundamental approach and adopt a practical point of view: we study the robustness and the erasure cost evolution of a logic gate when it is used repeatedly. In other words, we investigate on a memory response to successive use without letting the system relax to its initial equilibrium configuration in between.

Our experiments are based on an underdamped oscillator confined in a double-well potential, used as a model for a memory~\cite{Dago-2021,Dago-2022-JStat}. This system can be operated fast: an erasure can be reliably performed in just a few oscillation periods of the resonator. Its low dissipation then allows us to contain the overhead to Landauer's bound~\cite{Dago-2023-AdiabaticComputing,Dago-2022-PRL}. There are however two important counterparts to this low damping: the intrinsic relaxation time is large, and the heat exchanges with the environnement are reduced. The consequence is that after a fast erasure, which requires some energetic input (the erasure work, at least $W_{LB}$), the memory temperature raises: the system stays out of equilibrium for a long time. The logical outcome and energetic cost of \emph{successive} operations is therefore an open question that we tackle in the next sections. In the first one, we introduce our experimental setup and summarize the consequence of a single high speed, low dissipation erasure: a warming of the memory, transiently up to doubling its temperature. In section~\ref{sec:repeatederasure}, we present our repeated erasures procedure and analysis criteria. The experimental results are presented in section~\ref{sec:expresults}, and modeled in sections \ref{sec:toymodel} and \ref{sec:fullmodel}: first with a simple toy-model giving a simple reliability criterion, then with a more complete semi-quantitative model. We conclude the article with a discussion on the practical consequences and limitations of underdamped memories, and provide a cost map to optimize the choice of the memory characteristics with respect to the operation speed, reliability and power consumption.

\section{Single fast erasure in underdamped memories} \label{sec:singleerasure}

\begin{figure}[ht]
	\includegraphics[width=8.5cm]{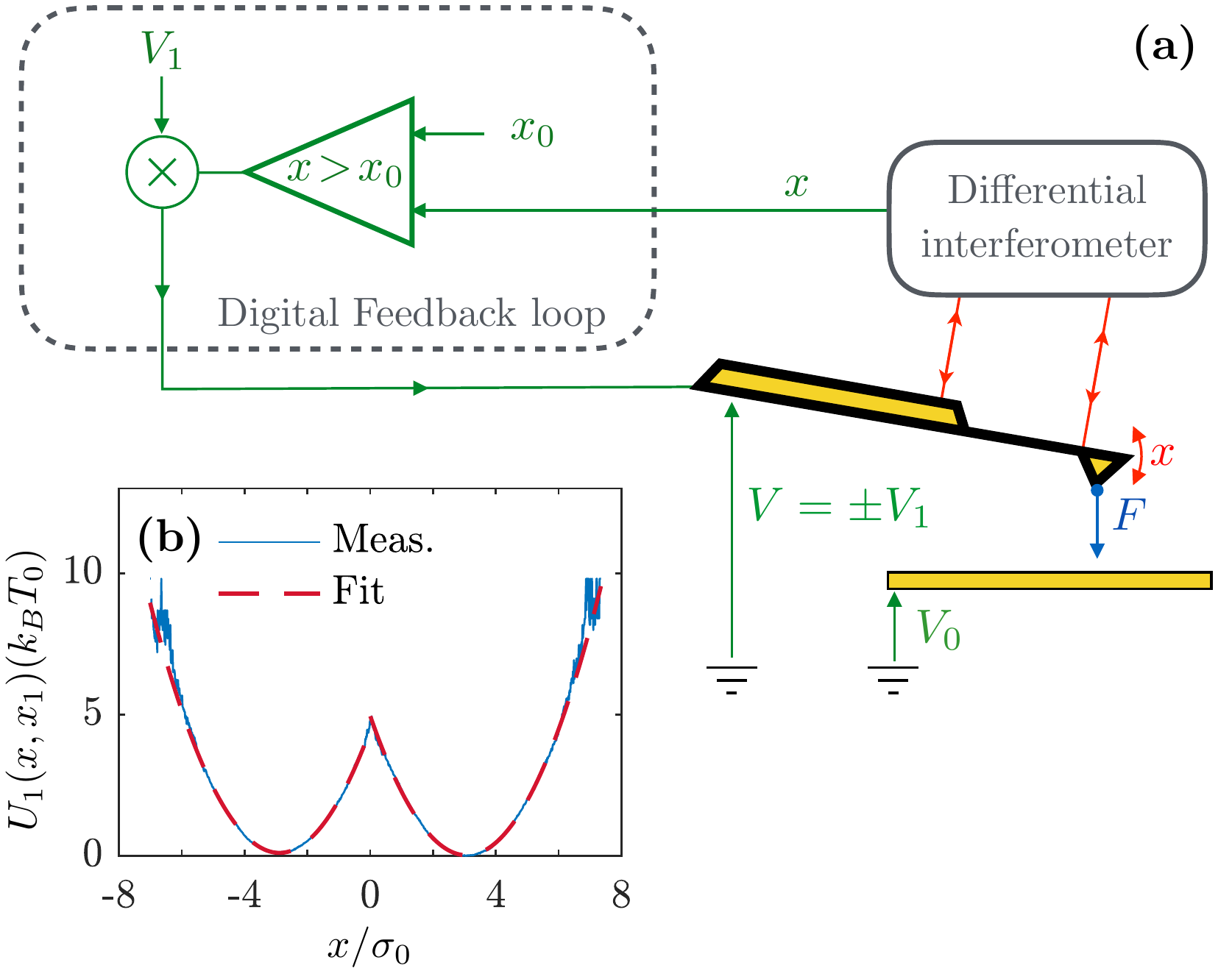}
	\caption{\textbf{(a) Sketch of the experiment.} A conductive cantilever (yellow) is used in vacuum as an underdamped harmonic oscillator. Its deflection $x$ is measured by a high resolution interferometer, and compared in a fast digital feedback loop to the user controlled threshold $x_0$. This feedback applies a voltage $\pm V_1$ to the cantilever depending on the sign of $x-x_0$. An electrostatic force between the cantilever and a facing electrode at voltage $V_0\gg V_1$ displaces the center of the harmonic well potential to $\pm x_1$, with $x_1$ tunable via the voltage $V_1$. When $x_0>x_1$ (respectively $x_0<x_1$), this results in a single well centered in $-x_1$ (respectively $+x_1$), and if $x_0=0$, in a virtual bi-parabolic potential energy $U_1(x,x_1)=\frac{1}{2}k(| x | -x_1)^2$. \textbf{(b) Double well potential.} From the statistical distribution of $x$ at equilibrium in the double well, we reconstruct using Boltzmann's distribution the effective potential energy felt by the oscillator. The bi-parabolic fit (dashed red) is excellent, and demonstrates a barrier of $\frac{1}{2}kx_1^2=5k_BT_0$ in this example. $x$ is normalized by its standard deviation $\sigma_0$ at equilibrium in a single well.}
	\label{setup}
\end{figure}

Before exploring the memory performance in response to a concrete use (successive use rather than independent single erasures), we summarize in this section the behavior of underdamped memories subject to a single fast erasure, in the framework of our previous explorations~\cite{Dago-2022-PRL,Dago-2023-AdiabaticComputing}. Our experiment is built around an underdamped micro-mechanical resonator (a cantilever in light vacuum at $\sim\SI{1}{mbar}$) characterized by its high quality factor $Q\sim 80-100$, effective mass $m$, natural stiffness $k$, leading to a resonance angular frequency of $\omega_0=\sqrt{k/m}=2\pi \times (\SI{1350}{Hz})$. The position $x$ of the oscillator is measured with a high precision differential interferometer~\cite{Bellon-2002,Paolino2013}, and due to thermal noise its variance at rest is $\sigma_0^2=\langle x^2 \rangle=k_BT_0/k\sim\SI{1}{nm^2}$, with $T_0$ the bath temperature. Thanks to a fast feedback loop and an electrostatic actuation, the oscillator can be operated in a double-well bi-quadratic potential, $U_1(x,x_1)= \frac{1}{2}k(| x | -x_1)^2$, with $x_1$ the user-controlled parameter tuning the barrier height~\cite{Dago-2021,Dago-2022-JStat}. This double well allows us to define a 1-bit information: its state is 0 (respectively state 1) if the system is confined in the left (respectively right) hand well of $U_1$. At rest, we use $x_1=X_1 \gtrsim 5 \sigma_0$, corresponding to an energetic barrier $\B=\frac{1}{2}kX_1^2=12.5k_BT_0$ high enough to secure the initial 1-bit information. A sketch of the setup and an example of double well are displayed on Fig.~\ref{setup}. Further details on the experimental setup and the validity of the virtual potential constructed by the feedback loop are discussed in Ref.~\onlinecite{Dago-2022-JStat}.

\begin{figure}[t]
\includegraphics{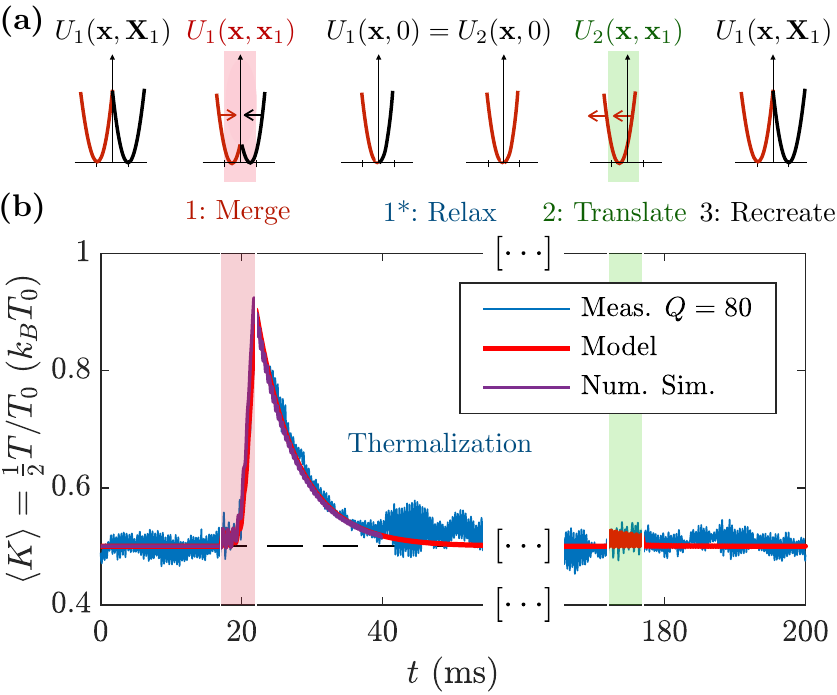}
\caption{\textbf{Single erasure protocol. (a) Schematic view.} Snippets of the potential energy during the erasure protocol. We start at equilibrium in a double well potential $U_1(\x,\x_1=\X_1)$, then proceed with: step 1 [Merge] to merge the two wells together into a single one well centered in $0$; step 1* [Relax] allowing the system to equilibrate in the single well; step 2 [Translate] to move the single well $U_2(\x,\x_1)$ to the position $-\X_1$ of state 0; finally step 3 [Recreate] to get the initial potential back by recreating the second well in position $+\X_1$ .
\textbf{(b) Kinetic energy during a fast erasure}. Step 1 [Merge] in red background lasts $\tau=\SI{5}{ms}$ (with $\X_1=5$, corresponding to $\mathbf v_1=0.12$) and results in a strong temperature rise visible on the kinetic energy profile: $\langle K \rangle$ culminates at $\SI{0.92}{k_BT_0}$, close to the adiabatic limit $K_a=k_BT_0$ \cite{Dago-2023-AdiabaticComputing}. At the end of the compression step, the system thermalizes with the surrounding bath in $\tau_{relax}\sim\SI{20}{ms}$ during step 1* [Relax] so that the kinetic energy relaxes to its equilibrium value $K_{\textrm{eq}} = \frac{1}{2} k_B T_0$. Then, the translational motion of duration $\tau$ (step 2 [Translate] in green background) only produces tiny oscillations. The model without any tunable parameters (red) nicely matches the experimental curve (blue) averaged from $1000$ trajectories and the simulation results for step 1 (purple) obtained from $10^5$ simulated trajectories.}
\label{Kadiab}
\end{figure}

The basic erasure procedure is similar to the standard approach used in previous stochastic thermodynamics realizations~\cite{Berut2012, Berut2015, orl12, Bech2014, Gavrilov_EPL_2016, Finite_time_2020, Hong_nano_2016, mar16}: lower the barrier, tilt the potential towards the reset state, raise the barrier. In our case, it corresponds to the following steps: 
\begin{enumerate}
\itemsep0.3em 
\item {[}Merge]: $x_1$ is decreased from $X_1$ to $0$ in a time $\tau$, corresponding to the dimensionless speed $\mathbf v_1=X_1/(\sigma_0\omega_0\tau)$ of the center of the wells of $U_1(x,x_1)$. This results in merging the two wells into a single one, effectively compressing the phase space along the spatial coordinate $x$. This step implies a warming of the underdamped system for high speeds or low damping: the heat flux with the bath is not efficient enough to compensate the compression work influx~\cite{Dago-2022-PRL,Dago-2023-AdiabaticComputing}. 
\item[1*.] {[}Relax]: during this optional step, the system is left at rest in the single well in order to equilibrate it with the thermal bath after the wells merging.
\item {[}Translate]: the single well described by the potential $U_2(x,x_1)= \frac{1}{2}k(x \pm x_1)^2$ is translated to the final position $\mp X_1$ by ramping $x_1$ from 0 to $X_1$ at the same speed $\mathbf v_1$ as in the previous step.
\item {[}Recreate]: the potential is switched back to the initial one $U_1^0=U_1(x,X_1)$, thus recreating an empty well on the opposite side of the reset position.
\end{enumerate}
Starting from equilibrium, this procedure (summarized in Fig.~\ref{Kadiab}(a) has a 100\% success rate: it always drives the system to the target state independently of its initial one.

We define the system internal temperature $T$ via the average value of the kinetic energy $\langle K \rangle = \langle \frac{1}{2}mv^2 \rangle = \frac{1}{2}k_BT$ (see \methods~\ref{sec:KinT}). When the memory is at equilibrium with the bath at temperature $T_0$, the equipartition imposes $\langle K \rangle_{eq} =\frac{1}{2}k_BT_0$, so that as expected $T=T_0$. In Fig.~\ref{Kadiab}b we observe the time evolution of $\langle K \rangle$ during an erasure performed at $\mathbf v_1=0.12$. We observe on the experimental and numerical simulation results that the temperature increases as expected during step 1, followed by a slow relaxation to $T_0$ during step 1*. Then, step 2 fast translation triggers tiny transient oscillations. The temperature profile is successfully modeled following Ref.~\onlinecite{Dago-2022-PRL}. Let us emphasize again that this warming of the memory is due to the high quality factor and erasure speed, as both result in inefficient heat exchanges with the bath. It saturates at the adiabatic limit $T_a=2T_0$ when the heat exchanges are negligible during step 1~\cite{Dago-2023-AdiabaticComputing}. In Fig.~\ref{Kadiab}(b), for $\mathbf v_1=0.12$, the kinetic energy only approaches the adiabatic limit $K_a=\frac{1}{2} k_BT_a=k_BT_0$.

If step 1* [Relax] is long enough to let the system reach the equilibrium before step 2 [Translate], then one can perform successive and equivalent erasures, maintaining a constant operation cost. Nevertheless, one may wonder on a practical point of view, what happens if the erasures are repeated without waiting for equilibrium between each steps. Such a procedure would get rid of the long relaxation times (for example $\SI{20}{ms}$ in Fig.~\ref{Kadiab}(b) and significantly shorten the process. 

Let us finally point out that at high damping (overdamped case), the instantaneous thermalization allows one to sequence erasures without consequences on the thermodynamics, but faster erasure requires a huge energetic cost (to compensate the viscosity). That is why, to optimize the information processing speed and cost it is worth considering the very low damping regime. In this context, we detail in the following how the erasure cost is impacted by the removal of equilibration steps and by the repetition. In the light of previous findings, for fast erasures we expect that the temperature should increase continuously in average: it rises during the compression without having enough time to relax to $T_0$ before the next protocol. The temperature increments could nevertheless saturate at one point, if the energy surplus gained at each compression is compensated by the heat exchanges with the bath.

\begin{figure}[b]
	\centering
	\includegraphics{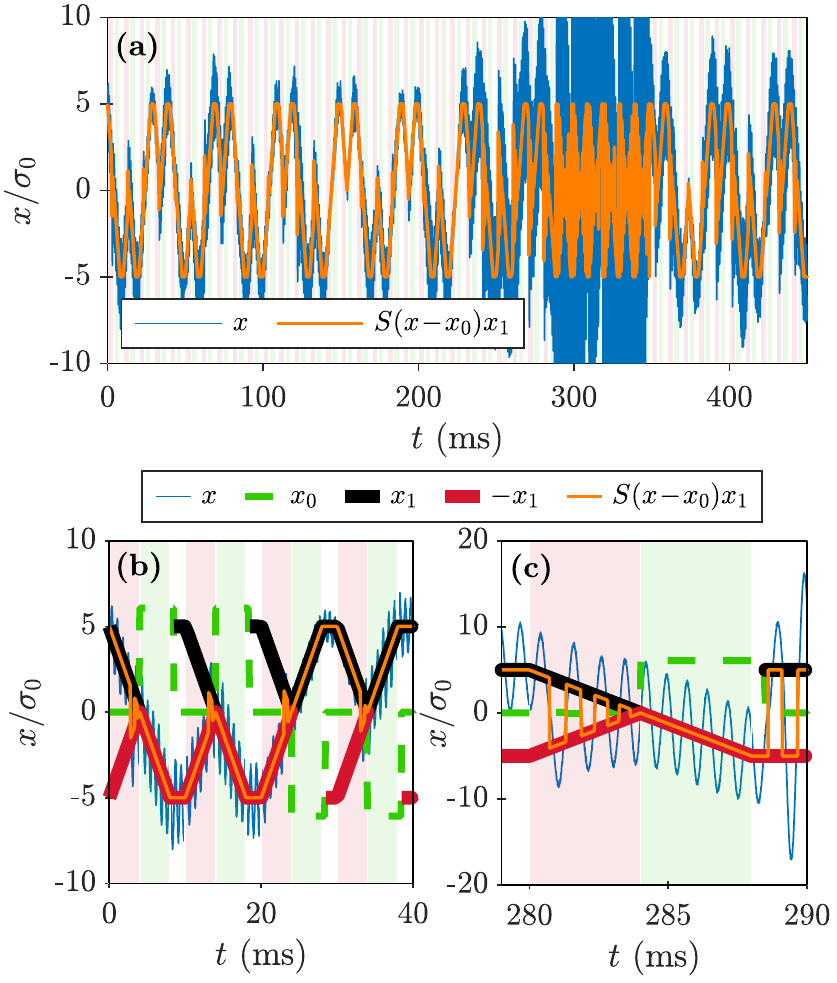}
	\caption{ \textbf{(a) Protocol of 45 repeated erasures.} The [Merge] and [Translate] steps (duration $\tau=\SI{4}{ms}$) and the [Recreate] one (duration $\tau_r=\SI{2}{ms}$) of each erasure are respectively highlighted in red, green and white background. During step 1 switches occur in the double well potential, whereas during step 2 the cantilever is driven towards the target state in a single well. One example of trajectory $x(t)$ is plotted in blue, it evolves at all times into the well centered in $S(x-x_0)x_1$ (orange), where $S(\cdot)=\pm 1$ is the sign function and $x_0(t)$ is the threshold imposed by the protocol: $x_0=0$ for the double well $U_1$ (step 1 and end of step 3), $x_0=\pm 6\sigma_0$ for single well $U_2$ targeting $\mp X_1$ (step 2 and beginning of step 3). During the first $\SI{280}{ms}$ of this example, 22 erasures are performed successfully. Afterwards, the operation fails several times as the system energy is too high. \textbf{(b) Zoom on the 4 first erasures, all successful.} The cycle covers all combinations $(0,1) \rightarrow 0$ and $(0,1) \rightarrow 1$: reset to 0 for the first two, reset to 1 for the next two. Each erasure starts in the double well $U_1^0$ ($x_0=0$, $x_1=X_1=5\sigma_0$), merged into a single well during step 1 ($x_0=0$, $x_1\rightarrow 0$), then driven towards the target state during step 2 ($x_0=\pm 6\sigma_0$ depending on the target state, $x_1\rightarrow X_1$), and $U_1^0$ is finally recreated during step 3. We evaluate the success of the erasure during the last $\SI{1.5}{ms}$ of step 3. \textbf{(c) Zoom on the first failure at $\SI{280}{ms}$}. During the free evolution in the final double well $x(t)$ escapes the target state (here state 0): the erasure fails.}
	\label{successive_erasures}
\end{figure}

\section{Repeated erasures protocol and reliability criteria} \label{sec:repeatederasure}

To explore the sustainability of repeated operations in a small amount of time, we perform 45 successive erasures with no step 1* [Relax], as plotted in Fig.~\ref{successive_erasures}(a).

The pattern is then the following: start with step 1 [Merge], duration $\tau$; immediately follow with step 2 [Translate], same duration $\tau$ (alternatively targeting state 0 or 1); and end by step 3 [Recreate], duration $\tau_r=\SI{2}{ms}$, with a free evolution of $\SI{0.5}{ms}$ in the single well followed by $\SI{1.5}{ms}$ into $U_1^0$ to evaluate the success of the erasure; finally cycle. As alluded to for step 2, to probe equiprobably all resetting configurations, we tackle the two procedures $(0,1) \rightarrow 0$ and $(0,1) \rightarrow 1$, that reset the memory to either state 0 or state 1 respectively. As we sequence the operations, the initial state of one erasure corresponds to the final state of the previous one: to cover equivalently the whole $(0,1)$ initial state space, we choose to cycle on 4 erasures \raisebox{-0.3pt}{\includegraphics{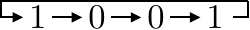}}. The choice of the final state has no impact on the thermodynamics because our erasing procedure is symmetric: any configuration therefore contributes evenly to the statistics.

Fig.~\ref{successive_erasures}(b) shows a cycle of four successful operations, the first two erasing to state 0 and the last two to state 1. In contrast, we plot on Fig.~\ref{successive_erasures}(c) an erasure failing: the system ends up with more energy than the barrier $\B$ so that the final state (state 0 here) is not secured. As a consequence, during the $\SI{1.5}{ms}$ free evolution in $U_1^0$ of step 3 before the next repetition starts, the system switches between state 0 and state 1. This erasure is classified as a failure: if we wait for the system to relax afterwards, it will end randomly in state 0 or 1 instead of the prescribed state. When the procedure fails once, we do not consider all the subsequent erasures since the initial state is undetermined. 

Fig.~\ref{successive_erasures}(c) is a zoom on the first failure of the protocol plotted in Fig.~\ref{successive_erasures}(a). We see indeed the deflection excursion growing progressively, until the systems no longer ends in a secured final state after $i=22$ erasures. We note $N_i$ the number of trajectories ensuring a successful outcome of the erasure $i$. We deduce from the $N_0=2000$ protocols the average success rate at each repetition: $R^{\textrm{s}}_i=N_i/N_0$. When an erasure is a success we compute the stochastic work and heat, and deduce the average values from the $N_i$ trajectories. 

\section{Experimental results} \label{sec:expresults}

\begin{figure}
	\centering
	\includegraphics{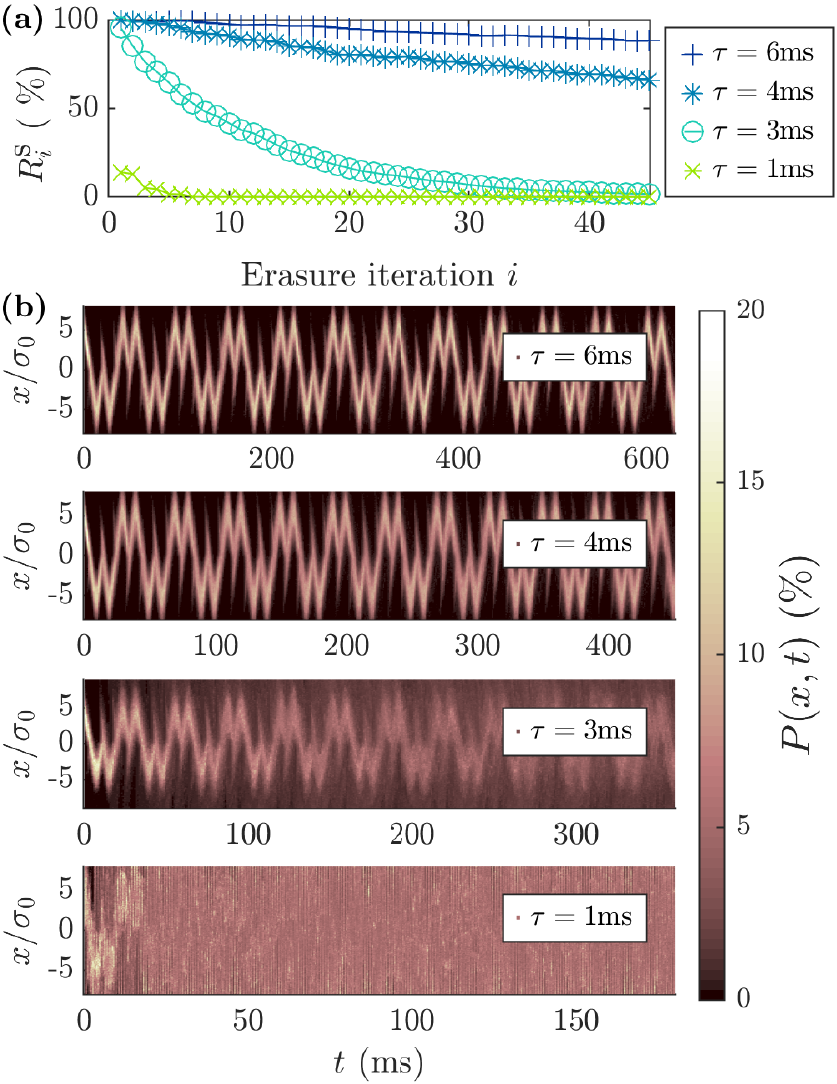}
	\caption{\textbf{(a) Success of repeated erasures for different operation speeds.} Success rate of the iteration $i$ of the 45 repeated erasures: $R^{\textrm{s}}_i=N_i/N_0$, computed from $N_0=2000$ procedures at $\tau=\SI{6}{ms}$, $\tau=\SI{4}{ms}$, $\tau=\SI{3}{ms}$ and $\tau=\SI{1}{ms}$. An erasure is classified as a success as long as the cantilever stays in the desired final state during the $\SI{1.5}{ms}$ free evolution in $U_1^0$ at the end of step 3. We distinguish the speed region C ($\mathbf v_1 < 0.2$, corresponding here to $\tau=\SI{4}{ms}$ and $\SI{6}{ms}$) resulting for the major part in a protocol success ($R^{\textrm{s}}_{45}>70\%$), from the region D ($\mathbf v_1 >0.2$, corresponding to $\tau=\SI{1}{ms}$ and $\SI{3}{ms}$) in which the memory fails to repeat successfully the operation ($R^{\textrm{s}}_{45}<2\%$). \textbf{(b) Measured probability density $P(x,t)$} inferred from $N_0=2000$ trajectories for different speeds. $P(x,t)$ is normalized at each time $t$. The blurring of the probability density is consistent with the success rate: when the excursion goes huge, the trajectories escape the driving and the information is lost.}
	\label{Pzt_success}
\end{figure}

The goal of the experiment is to explore the robustness of the memory to repeated erasures depending on the speed imposed to perform one operation. We compare the responses for $\tau=6$, $4$, $2$ and $\SI{1}{ms}$, tackling the high speed limit of our setup. It corresponds to $\mathbf v_1 = 0.1$, $0.15$, $0.3$ and $0.6$, so that the last dataset allows only one oscillation for each step. Let us point out that the total erasure duration is worth $2\tau+\tau_r$ with $\tau_r$ fixed to $\SI{2}{ms}$.

For different speeds we compute the success rate $R^{\textrm{s}}_i$ of the erasure $i$, the probability density in position $P(x,t)$, the average kinetic energy evolution $\langle K \rangle$, and the average work $\langle \W \rangle$ required for each successful erasure. Let us first tackle the success rate and the probability density plotted in Fig.~\ref{Pzt_success}. Without surprise, the faster the information is processed, the less reliable the operation becomes. Indeed for $\tau=\SI{6}{ms}$ and $\tau=\SI{4}{ms}$ the success rate after 45 repeated erasures stays above $70 \%$; meanwhile for $\tau=\SI{3}{ms}$ and $\tau=\SI{1}{ms}$ it collapses after a few erasures and the probability to complete the whole protocol is null. These two driving speed regimes ($\mathbf v_1< 0.2$ called region C for Converging and $\mathbf v_1\geq 0.2$ called region D for Diverging) are also visible on the probability density in Fig.~\ref{Pzt_success}(b). Within the speed region C ($\tau=\SI{6}{ms}$ and $\SI{4}{ms}$) the trajectories remain mostly contained by the double well barrier height and the information isn't lost ($R^{\textrm{s}}_{45}>70\%$). On the contrary, very fast procedures (speed region D, $\tau=\SI{3}{ms}$ and $\SI{1}{ms}$) result in the blurring of $P(x,t)$, because the oscillator overcomes more and more often the double well barrier: the reset systematically fails at the end of the protocol ($R^{\textrm{s}}_{45}<2\%$).

\begin{figure}
	\centering
	\includegraphics{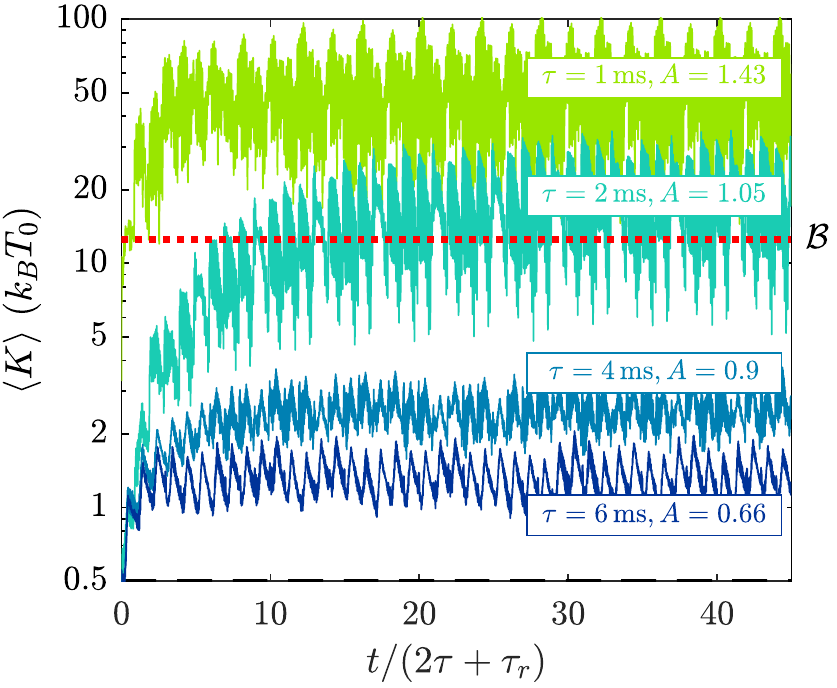}
	\caption{\textbf{Average kinetic energy during 45 successive erasures.} For speed region C ($\tau=4$ and $\SI{6}{ms}$, $\mathbf v_1 < 0.2$), $\langle K \rangle$ starts from its equilibrium value ($\frac{1}{2} k_B T_0$ in dashed line) and nearly doubles during the first successive adiabatic compressions. The thermalization afterward is only partial and insufficient to stabilize the temperature for the first iterations. Eventually the kinetic energy converges to a plateau after a couple of erasures; the higher the speed the higher the saturation value. On the other hand, for speed region D ($\tau=1$ and $\SI{2}{ms}$, $\mathbf v_1 > 0.2$) the kinetic energy strongly increases and overreaches the barrier height (dotted blue): the thermalization doesn't balance the compression warming anymore. Moreover, if the operation fails, the system ends in the wrong well and takes an energy kick when the potential $U_1^0$ is rebuilt. As a consequence, a runaway occurs because more failures result in energy peaks and energy rise leads to more failures. The simple model successfully predicts these two regimes with the indicator $A$ (computed with Eq.~\ref{A} assuming $Q=90$): $A<1$ in speed region C (Convergence to a plateau below $\B$) and $A>1$ in speed region D (Divergence beyond $\B$).}
	\label{K_all_DVG}
\end{figure}

\begin{figure}
	\centering
	\includegraphics{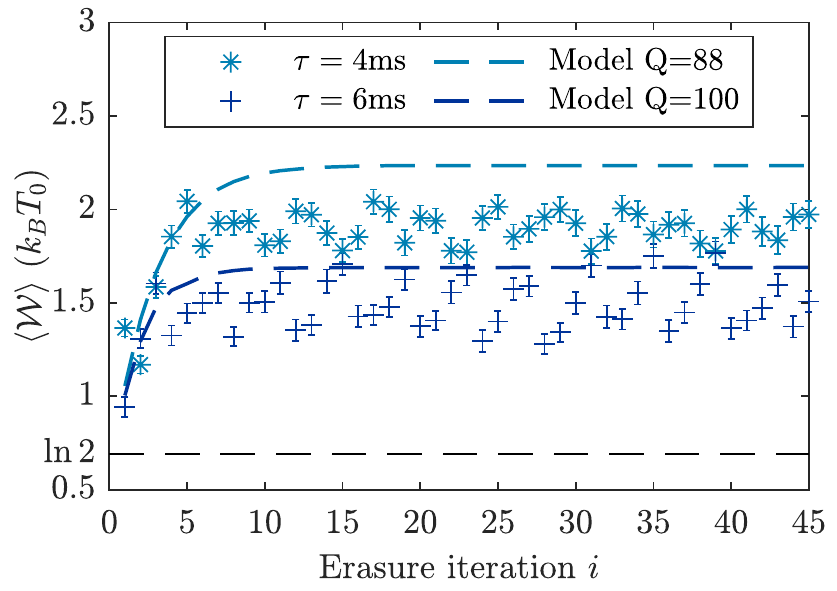}
	\caption{\textbf{Average work per erasure during 45 successive operations.} $\langle \W \rangle$ of erasure $i$ is inferred from the $N_i$ successful trajectories, for the 2 erasure speeds in region C allowing enough successful erasures ($\tau=\SI{4}{ms}$ and $\tau=\SI{6}{ms}$). After a couple of repetition the average work reaches a plateau depending on $\tau$: $\W_{\textrm{sat}}(\tau=\SI{6}{ms})=\SI{1.5}{ k_BT_0}$, and $\W_{\textrm{sat}}(\tau=\SI{4}{ms})=\SI{1.9}{k_BT_0}$. The model (dashed line) successfully predicts the converging behavior. It is in reasonable agreement with the experimental result considering the approximations made and the fact that near the region boundary it is very sensitive to the calibration parameters. The quality factor $Q$ used for the model is the same as the one tuned to match the kinetic energy profile in Figs.~~\ref{K_model_4-6ms}.}
	\label{W_comp}
\end{figure}

The success rate can be explained by the temperature profile of the memory visible through the average kinetic energy plotted in Fig.~\ref{K_all_DVG}. During the first repetitions, the temperature nearly doubles at each compression, and decreases afterward without fully thermalizing. For $\tau=\SI{6}{ms}$ and $\tau=\SI{4}{ms}$ it finally stops increasing by step and reaches a permanent regime below the barrier allowing a secure encoding of the information. On the other hand, for $\tau=\SI{3}{ms}$ and $\tau=\SI{1}{ms}$ the kinetic energy skyrockets and exceeds the energy barrier. We recover through the temperature behavior the two speed regions identified when analyzing the success rate. 

When the erasure succeeds, it is also interesting to quantify the work required on average. Indeed at the same speed, since the memory is hotter after a repeated use, we expect the work to be higher than the one required for a single erasure studied in Ref.~\onlinecite{Dago-2021}. As we tackle the erasure work, we restrict the study to region C where there are enough successful operations to compute properly the erasure cost for the 45 iterations: the operation cost for speeds $\tau=\SI{6}{ms}$ and $\tau=\SI{4}{ms}$ is displayed in Fig.~\ref{W_comp}. It highlights that not only the failure rate increases with the speed, but also the work required to process the information. Indeed after a quick transient, the work reaches a plateau whose value grows with the speed of the process. We detail in the next sections first a simple model that helps understanding and predicting the energy behavior, and second a more complete description to provide semi-quantitative results.

\section{Simple Model} \label{sec:toymodel}

The goal of this section is to propose a very simple model to grasp the behavior of the memory in response to successive use, and in particular to understand the two speed regimes observed experimentally. Indeed within region D the kinetic energy widely exceeds the barrier, leading to the systematic failure of the protocol after several repetitions. On the other hand erasures in region C has a kinetic energy converging below the barrier and a good success rate. This behavior can be explained by the balance between the warming during the compression and the heat continuously released into the bath. Depending on the relative importance of this two opposite phenomena appears or not a saturation temperature allowing the two to compensate each-other.

\begin{figure}
	\centering
	\includegraphics[width=\columnwidth]{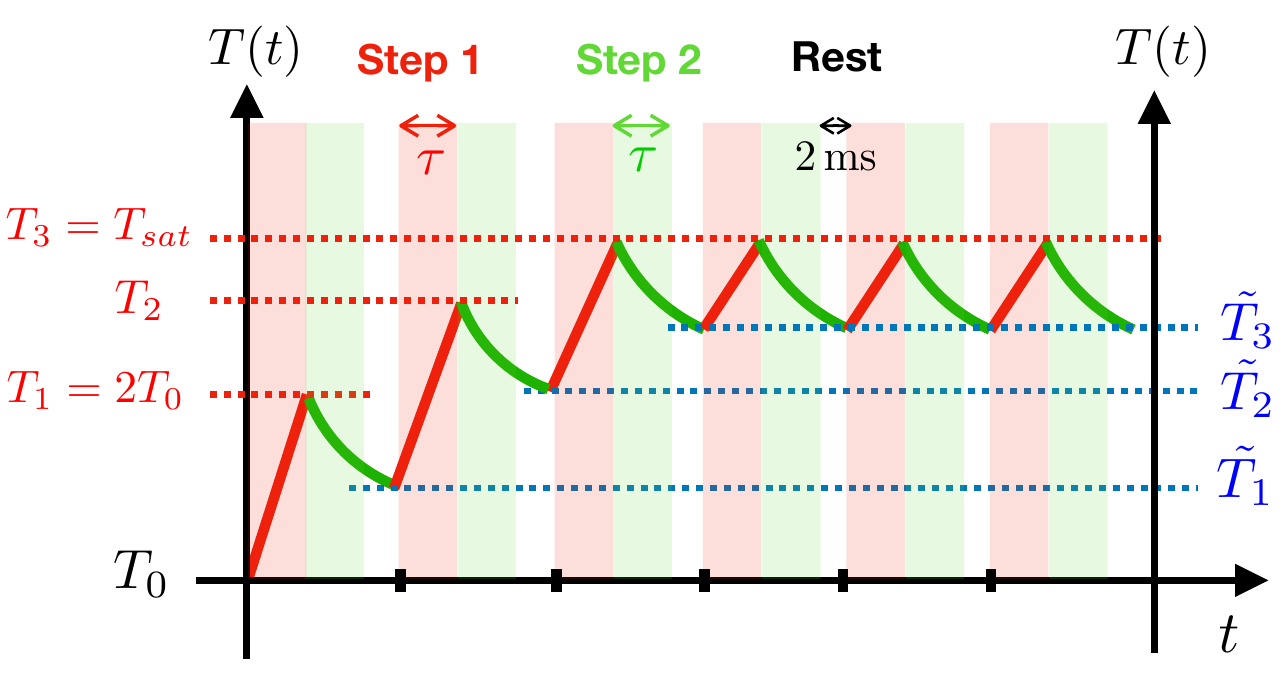}
	\caption{ \textbf{Schematic description of the simple model}. We decompose the protocol in successive steps 1 (red segments, duration $\tau$), followed by steps 2 and 3 (green segments, duration $\tau_r=\SI{2}{ms}+\tau$). For each erasure $i$ we call $T_i$ the temperature after step 1 and $\tilde{T}_i$ the temperature after the relaxation. After some repetitions the temperature can either converge and saturate to a permanent regime, or diverge and exceed the barrier.}
	\label{simple_model}
\end{figure}

From the temperature rise perspective, the erasures of the repeated procedure are decomposed into the compression, step 1 lasting $\tau$; and the thermalization, steps 2 and 3 lasting $\tau+\tau_r$. We introduce the following notations (illustrated in Fig.~\ref{simple_model}): the maximum temperature of erasure $i$ reached at the end of the step 1 is $T_i=\alpha_i T_0$, and the temperature at the end of the thermalization is $\tilde{T}_i$. The initial temperature is $\tilde{T_0}=T_0$.

To build the simple model we start with the energetic balance of the system: 
\begin{equation}
\frac{d \langle E \rangle}{dt} = \frac{d \langle \W \rangle}{dt} - \frac{d\langle \Q \rangle}{dt}
\label{eq_balance_cycle}
\end{equation}
Several assumptions and approximations justified by the high speed and quality factor are made to simplify the description:
\begin{enumerate}[label=(\roman*)]
\itemsep0em 
 \item During step $i+1$ starting at temperature $\tilde{T}_{i}$, the work expression in the adiabatic limit holds~\cite{Dago-2023-AdiabaticComputing}: $\W_a=k_B\tilde{T}_{i}$.
 \item Deterministic contributions ($K_D$ and $U_D$) are neglected (see \methods~\ref{deterministic_terms} for their definition), hence step 2 involves no work.
 \item The derivatives in Eq.~\ref{eq_balance_cycle} are taken at first order.
 \item Equipartition holds at the end of each steps, so that $\langle E \rangle=k_B T$.
 \end{enumerate} 

Hypotheses (ii) and (iv) allow us to simplify Eq.~\ref{eq_balance_cycle} during the thermalization into: 
\begin{align}
\frac{dT}{dt} = \frac{1}{k_B}\frac{d \langle E \rangle}{dt} &=- \frac{1}{k_B}\frac{d\langle \Q \rangle}{dt} =- \frac{\omega_0}{Q} (T(t)-T_0). \label{eq_balance_relax}
\end{align}
The last equality of previous equation stems from the general expression of heat in underdamped stochastic thermodynamics as long there is no deterministic kinetic energy in the system~\cite{Dago-2022-PRL}. From Eq.~\ref{eq_balance_relax}, we deduce that the temperature initially at $T_i$ relaxes exponentially towards $T_0$ during $\tau+\tau_r$ (green segments in Fig.~\ref{simple_model}), so that:
\begin{align}
\tilde{T}_i&=T_0+(T_i-T_0) e^{-\frac{(\tau_r+\tau) \omega_0}{Q}}\\
&=(1-r)T_0+r T_i, \textrm{ with } r=e^{-\frac{(\tau_r+\tau) \omega_0}{Q}} \\
&=[ 1+r(\alpha_i -1) ]T_0.
\end{align}

We now address the erasures step 1 (red segments in Fig.~\ref{simple_model}) using hypotheses (i), (iii) and (iv) to rewrite the energy balance (Eq.~\ref{eq_balance_cycle}) as:
\begin{align}
k_B \frac{T_{i+1}-\tilde{T}_{i}}{\tau}&=\frac{k_B \tilde{T}_i}{\tau} - \frac{\omega_0}{Q}k_B (\tilde{T}_{i}-T_0). \label{eq-EnergyBalanceStep1}
\end{align}
In this expression, we used as the relevant heat derivative its initial point ($\frac{d\langle \Q \rangle}{dt}(t)\simeq k_B\frac{\omega_0}{Q} (\tilde{T}_{i}-T_0)$). Expressing all temperatures with $\alpha_i$, we get
\begin{align}
\alpha_{i+1}&=r(2-\frac{\omega_0 \tau}{Q})(\alpha_{i}-1)+2
\end{align}
We recognize a geometric serie: $\alpha_{i+1}=A\times \alpha_{i} +B$, with $\alpha_0=1$, $B=2-A$ and: 
\begin{align}
A= e^{-\frac{(\tau_r+\tau) \omega_0}{Q}}(2-\frac{\omega_0 \tau}{Q}) \label{A}. 
\end{align}
All in all the model, exhibits two regimes: if $A<1$ the warming and thermalization compensates after some iteration, so that the temperature converges to $T_{sat}=(1+\frac{1}{1-A})T_0$; and if $A>1$, the heat exchange is inefficient to compensate the energy influx from the successive compressions and the temperature diverges. The parameter $A$ controlling the convergence, decreases with $\tau$ and increases with $Q$: these dependences fit with the experimental observations. We apply the simple model (assuming $Q=90$) and compute $A$ with Eq.~\ref{A} for the different experimental durations in Fig.~\ref{K_all_DVG}: the model successfully predicts the booming of the energy in region D. 

\begin{figure*}[t]
	\includegraphics{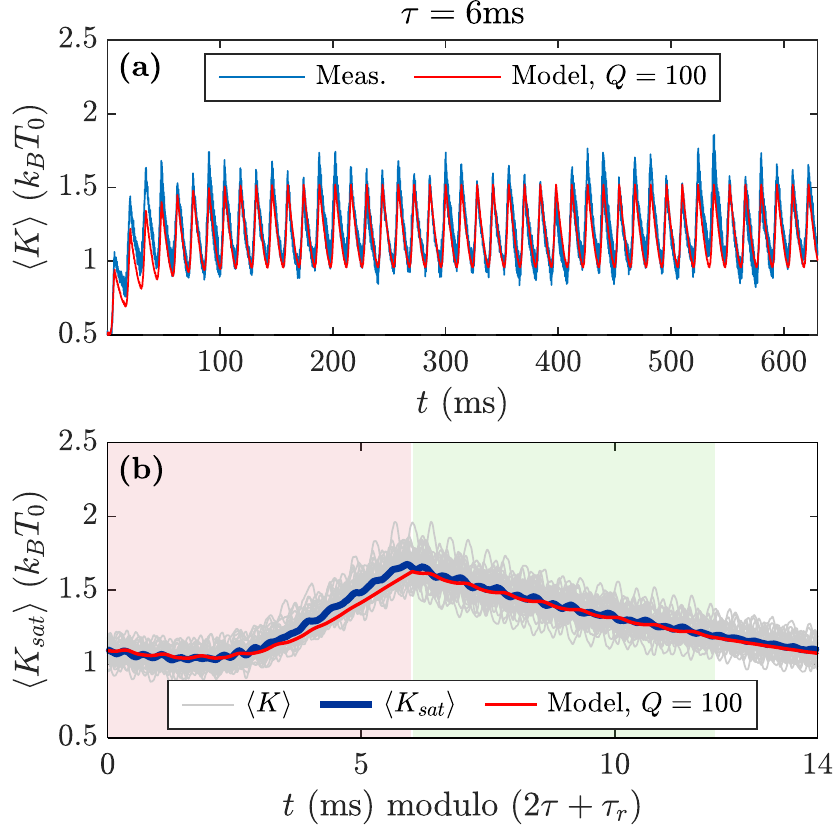} \hfill \includegraphics{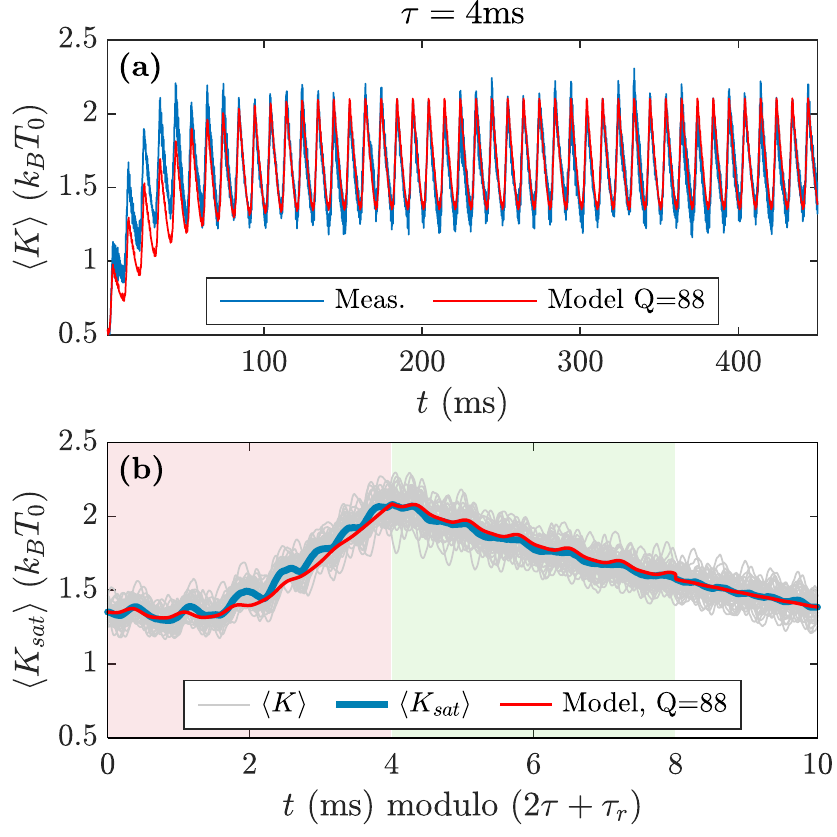}
	\caption{\textbf{Kinetic energy evolution for 45 repeated erasures}. The left (respectively right) panels correspond to a duration $\tau=\SI{6}{ms}$ (respectively $\tau=\SI{4}{ms}$). \textbf{(a) Average kinetic energy.} $\langle K \rangle$ (blue) is inferred from the $N_i$ successful trajectories of the erasure $i$ (as we are in speed region C, $N_i\sim N_0$), and plotted during the whole protocol. Initially at the equilibrium value ($\frac{1}{2}k_BT_0$), $\langle K \rangle$ nearly doubles during the first 3 to 5 compressions without fully thermalizing in-between, and eventually reaches a plateau: around $K_{sat}=1.3k_BT_0$ for $\tau=\SI{6}{ms}$, and around $K_{sat}=1.8k_BT_0$ for $\tau=\SI{4}{ms}$. The quantitative model (red) is in very good agreement with the experimental results with no adjustable parameters except for a tiny adjustment of the quality factor. \textbf{(b) Saturation profile.} When the permanent regime is established, $\langle K \rangle$ follows a repeated pattern every $2\tau+ \tau_r$: these similar profiles are superimposed in grey lines. The saturation curve $\langle K_{sat} \rangle$ (blue) is the average of the permanent regime profiles of the last 40 operations. The system first continues to relax from $1.1k_BT_0$ (left) or $1.4k_BT_0$ (right) at the beginning of step 1 (transient oscillations appear due to the translational motion), until the two wells get close enough and the compression actually starts, resulting in the temperature rise. During step 2 (green background) the system thermalizes with again transient oscillations, and keep on relaxing during the final $\tau_r=\SI{2}{ms}$ rest. The quantitive model (red) nicely matches the experimental curves: it consists in the theoretical model of Landauer's erasure (SE Model)~\cite{Dago-2022-PRL} using as initial kinetic temperature the experimental value $T_{sat}=2.2T_0$ (left) or $2.8T_0$ (right) measured on the permanent regime profile.}	
	\label{K_model_4-6ms}
\end{figure*}

As a conclusion to this section, this simple model includes many approximation but turns out to be enough to recover the speed region C and D corresponding to a converging or diverging evolution of the energy (respectively $A<1$ and $A>1$). The frontier $A=1$ corresponds to $\tau=\SI{3.34}{ms}$ which is again perfectly compatible with the experimental results. Nevertheless in the very fast and very slow limits, most of the assumptions may stop being relevant. In particular, the systems doesn't actually diverges when $A>1$ as predicted by the model but reaches a very high plateau. Indeed, if the system's energy broadly exceeds the barrier height, the potential driving protocol impact on the system's behavior becomes negligible, therefore making the above model meaningless. In particular the peaks observed in the permanent kinetic energy profile for region D in Fig.~\ref{K_all_DVG} no longer comes from the compression, but from the energy kicks given to the system when the barrier is restored to $x_0=0$ if the cantilever ended up in the wrong well.

\section{Quantitative Model} \label{sec:fullmodel}

In this section we propose a more detailed and complete model designed to quantitatively predict the system behavior in the converging region C. Indeed now that we have identified the speed interval allowed to successfully process repeated erasures, the point is to quantitively estimate the corresponding energetic cost and temperature evolution. In all the following we restrict the study to speed region C, and consider only successful erasures: in particular the average experimental kinetic energy $\langle K \rangle$ is now inferred from the successful operations only. 

At the basis of the quantitative description is the model developed in Ref~\onlinecite{Dago-2022-PRL} to describe Landauer's fast erasures called here SE (Single Erasure) Model. It has proven reliable to describe a single erasure starting at equilibrium and including an equilibration step 1* between step 1 and 2, as illustrated by the excellent agreement with the experimental data in Fig~\ref{Kadiab}. We adapt the SE model by removing the equilibration step: while the thermal contribution relaxes from step 1, there are transient oscillations due to the translational motion deterministic contribution. Doing so we successfully describe the first erasure and obtain the final temperature $\tilde{T}_1$. The strategy is then to use the model with a different initial condition: the initial temperature is no more set to $T_0$ but to $\tilde{T}_1$. All in all, the quantitative model of Repeated Erasure (RE model) consists in applying the SE model successively starting each time with the final temperature $\tilde{T}_i$ as initial temperature for the next iteration. Fig.~\ref{K_model_4-6ms} compare the RE model (red curve) to the experimental data (blue curve) for $\tau=\SI{6}{ms}$ and $\tau=\SI{4}{ms}$. All parameters are taken from the experimental data ($\omega_0$, $\tau$, $\tau_r$ and $X_1$) except from the quality factor that is being tuned within the interval $80<Q<100$ to provide the best fit to the experimental curves. Indeed the RE model is quite sensitive to the value of $Q$ near the divergence, and the uncertainty on the quality factor is not negligible: it may drift slightly during experimental runs or change in between them due to small vacuum drift. 

The RE model also computes the average work required for the repeated use of the memory: the prediction plotted in dashed lines on Fig.~\ref{W_comp} is reasonable (taking the same parameters as the one for the kinetic energy profile). Hence, the operator can theoretically estimate the excess of work required to perform successive erasures compared to a single one, and the number of repetition before reaching a permanent regime. However, even though the RE model has proven effective, it has some limitations. The deterministic part of the kinetic energy and of the work is inferred from translational motions starting from equilibrium, whereas in reality the system is always out of equilibrium either during step 1 or step 2. Besides, the model is inefficient to predict the consequences of an operation failure on the energy divergence: it only describes successful erasures. 

\begin{figure}[b]
	\centering
	\includegraphics{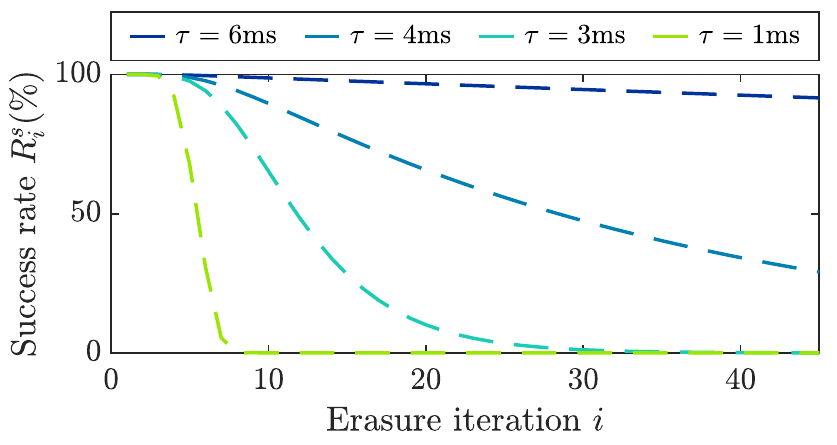}
	\caption{\textbf{Theoretical prediction of the erasure success rate for different speeds.} Assuming $Q=90$, we compute the escape rates $\Gamma(\B,\tilde T_k)$ using for $\tilde T_k$ the temperature theoretical profile (red in Fig.~\ref{K_model_4-6ms}), and deduce from Eq.~\ref{Rs} the success rate $R_i^s$ after $i$ erasures for the different $\tau$. Hence the model results in the probability of loosing the final information during the $\SI{1.5}{ms}$ free evolution in the final potential. As expected $R_i^s$ decreases with increasing speeds, and we identify region D in which the probability to successfully finish the whole protocol is zero.}
	\label{succes_model}
\end{figure}

Thanks to the theoretical knowledge of the temperature profile we are also able to approximate the success rate $R_i^s$ of $i$ successful repetitions of the operation. Indeed, the ratio $\B/(k_BT)$ ($\B$ being the barrier height) is all we need to compute the escape rate $\Gamma$ in the final double well potential~\cite{Dago-2023-ErratumJSTAT}:
\begin{equation}
\Gamma=\omega_0 \frac{k_B T}{\B} \frac{e^{-\B/(k_BT)} }{ \int_0^\infty d\epsilon \ e^{-\epsilon\B/(k_BT)} \left[\pi+ 2 \sin^{-1}{(\epsilon^{-\frac{1}{2}})}\right]}\label{EqGamma}.
\end{equation}
Note that in this expression, we extend the definition of the $\sin^{-1}$ function to arguments greater than one, with $\sin^{-1}(\epsilon)=\pi/2$ for $\epsilon>1$. If we assume that the temperature during the $\SI{1.5}{ms}$ final free evolution in $U_i^0$ is being worth $\tilde{T}_i$ (computed with the RE Model) we obtain the following success rate:
\begin{align}
R_i^s=\prod_{k=0}^{i-1} \big[1-\Gamma(\B,\tilde{T}_k)\tau_r\big].\label{Rs}
\end{align}
The result plotted on Fig.~\ref{succes_model} is qualitatively consistent with the experimental observations and quantify the consequence of the temperature rise on the success of the operation. Nevertheless, Eq.~\ref{EqGamma} accounts for the average escape time of a system at equilibrium in the initial well (at effective temperature $T$), while in reality there is a strong deterministic contribution just after step 2 that tends to push the system far from the barrier. This prediction of the reliability of erasure is thus quite conservative in general, but still provides a useful guideline for applications.

\section{Discussion and Conclusion}

Based on the previous studies of the energetic exchanges in an underdamped memory, we are able to grasp the consequences of its repeated use. Even though the low damping allows fast erasures at low energetic cost, the price to pay lies in the warming of the memory. As a consequence, if the memory is used several times straight after a previous operation without letting the system thermalize with its environment, the temperature rises by step. The thermal energy can then exceed the memory encoding barrier. The success of repeated operations therefore depends on the damping and the speed. On the first hand, the lower the damping, the longer the thermalization and the higher the compression warming: for a fixed speed, reducing the damping strengthens the temperature diverging behavior. On the other hand, the higher the speed, the higher the compression warming and the shorter the time allowed to thermalize: high speeds also favor the divergence.

\begin{figure}[tb]
\begin{center}
\includegraphics{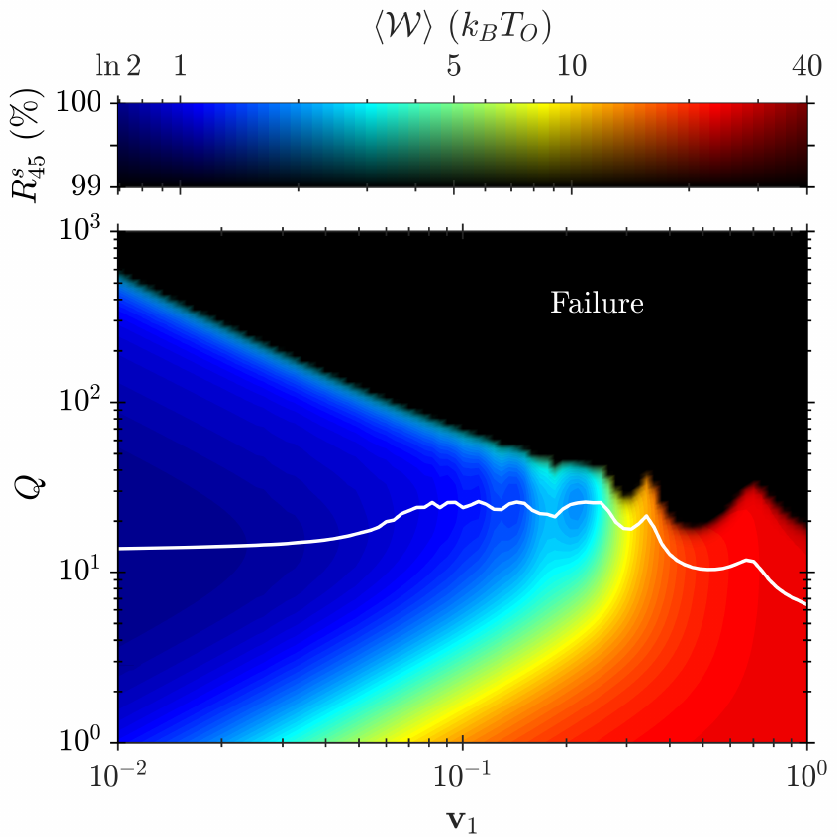}
\end{center}
 \caption{\textbf{Energetic cost and reliability of repeated erasures.} Both quantities are computed with the RE model in the permanent regime. The energetic cost is encoded by the colormap (red area corresponding to very consuming procedures and blue ones corresponding to frugal ones), and the shading gives the success rate of the operation (black area corresponding to a success rate below $99\%$). The optimal damping for sustainable and continuous 1-bit erasures is around $Q\sim10-20$ (white line, computed by the minimal work at each speed).}
 \label{map} 
\end{figure} 

We developed an efficient tool (the simple model) to predict the divergence of the energy, and therefore deduce the speed region which ensures a good success rate. Moreover, a more complete model can be used (the RE model) to quantitively estimate the energy and work evolution profile in response to repeated uses, and in particular the permanent regime reached after a few iterations. Fig.~\ref{map} summarizes the predictions of the RTE model for this permanent regime, with an hybrid map of energetic cost and operation reliability as a function of the two tunable parameters: quality factor $Q$ and erasure speed $\mathbf{v}_1$. From this map one can infer what is the optimal quality factor to minimize the erasure cost while maintaining a high success rate (black areas corresponding to a success rate below $99\%$). The final result, which could be used as a guideline for applications, is that optimal quality factors are $Q\sim 10-20$ for all speeds. Other protocols could be explored using the same theoretical approach, to further optimize erasure processes. It should be noted as well that a quality factor tunable on the fly during the protocol could reconcile the best of both worlds: high $Q$ during the compression to pay only the adiabatic erasure cost, followed by a low $Q$ during thermalization to cut the relaxation time and restore the initial equilibrium before the next operation. Such a strategy cannot be applied directly in our experiment: modifying the vacuum around the cantilever is slow. However, we could apply an extra damping force proportional to velocity using the feedback loop (which would also alter the effective thermostat temperature). In memories based on electronic devices, a switch could include an extra resistance in the circuit during thermalization, allowing fast tuning of $Q$.

As a conclusion, on a practical point of view, the underdamped regime thus appears to be an excellent choice to perform fast and repeated use of the memory at low cost. Indeed, the underdamped systems turns out to be quite robust to continuous information processing at high speed (only a few natural oscillation period per operations, here around 10 for the fastest reliable operations) at a stable and rather moderate cost (below $2k_BT_0$). Depending on the number of successive erasures one wants to perform and on the success rate required, the quality factor or the speed of the erasure have to be tuned to avoid divergences. 

\acknowledgments
This work has been financially supported by the Agence Nationale de la Recherche through grant ANR-18-CE30-0013. We thank J. Pereda for the initial programming of the digital feedback loop creating the virtual potential.

\section*{Methods}
\newcounter{AppendixEquation}
\makeatletter
\@addtoreset{equation}{AppendixEquation}
\makeatother

\newcounter{AppendixFigure}
\makeatletter
\@addtoreset{figure}{AppendixFigure}
\makeatother

\setcounter{section}{0}

\stepcounter{AppendixEquation}
\stepcounter{AppendixFigure}
\renewcommand{\thesection}{M\arabic{section}}
\renewcommand{\thetable}{M\arabic{table}}
\renewcommand{\thefigure}{M\arabic{figure}}
\renewcommand{\theequation}{M\arabic{equation}}

\section{Data}
The data that support the findings of this study are openly available in Zenodo~\cite{Dago-2023-DatasetAdvPR}.

\section{Underdamped stochastic thermodynamics}
\label{supp_thermo}
We consider a Brownian system of mass $m$ in a bath at temperature $T_0$ characterized by its position $x$ and speed $v$. Its dynamic into a potential energy $U(x)$ is described by the 1-dimension Langevin equation, 
 \begin{equation}
m \ddot{x} + \gamma \dot{x}=- \frac{dU}{dx}+\gamma \sqrt{D}\xi(t).
\label{Langevin1D}
 \end{equation}
The friction coefficient $\gamma$ of the environnement, the bath temperature $T_0$ and Boltzmann's constant $k_B$ define the diffusion constant through the Einstein relation: $D=k_B T_0/\gamma$. The thermal noise, $\xi(t)$, is a $\delta$-correlated white Gaussian noise:
 \begin{equation}
 \langle\xi(t)\xi(t+t')\rangle=2\delta(t').
 \label{xi}
 \end{equation}
  
We introduce the kinetic energy $\K=\frac{1}{2}m\dot x^2$. The equipartition gives the kinetic energy average value at equilibrium (as the potential does not depend on $v$):
 \begin{align}
\langle \K \rangle &=\frac{1}{2}k_BT_0.
\label{Eqpart_Ec}
\end{align}
As the total energy is worth $E=U+\K$, the energy balance equation writes: 
\begin{equation}
\frac{dK}{dt}+\frac{dU}{dt}=\frac{d\W}{dt}-\frac{d\Q}{dt},
\label{EqBalance}
 \end{equation} 
 with $\W$, the stochastic work defined by~\cite{sek10,sek66,Seifert_2012,Jarzynski_2011,Ciliberto_PRX,Dago-2022-PRL}:
\begin{align}
\frac{d\W}{dt}&=\frac{\partial U}{\partial x_1} \dot{x}_1, \label{dWsdt}
\end{align}
 and $\Q$ the stochastic heat defined by, 
 \begin{align}
\frac{d\Q}{dt}&\equiv-\frac{\partial U}{\partial x} \dot{x} -\frac{d\K}{dt}, \label{dQsdt} \\
&=\frac{\omega_0}{Q} (2 \langle \K \rangle-k_B T_0).
\label{Q}
\end{align}
Let us point out that the heat expression (Eq.~\ref{Q}) is completely general and doesn't depend on the potential shape or current transformations occurring in the system. It also highlights that for a large quality factor $Q$, the heat exchanges with the thermal bath are reduced. Finally, at equilibrium when the equipartition theorem prescribes $\langle \K \rangle=\frac{1}{2}k_B T_0$, there are in average no heat exchanges, as expected.

\section{Deterministic terms} \label{deterministic_terms}

The trajectory $x(t)$ in a moving well decomposes into the stochastic response to the thermal fluctuations, which vanishes on average, and the deterministic response: $x=x_{th}+x_D$, with $\langle x \rangle =x_D$. Similarly we define $\langle v \rangle=\dot x_D$. $x_D$ is the solution of the deterministic equation of motion: 
\begin{align}
\ddot x_D +\frac{\omega_0}{Q} \dot x_D - \frac{1}{m}\frac{\partial U}{\partial x}(x_D)=0. \label{Langevintrans1}
\end{align}
In a single quadratic well with a driving $x_1(t)$, Eq.~\ref{Langevintrans1} becomes: 
\begin{align}
\ddot x_D +\frac{\omega_0}{Q} \dot x_D + \omega_0^2 x_D&=\omega_0^2 x_1(t). \label{Langevintrans2}
\end{align}

As detailed in Ref.~\onlinecite{Dago-2022-PRL}, to best model the deterministic terms during the erasure protocol 
we express the deterministic work, kinetic and potential energies by: 
\begin{subequations} \label{eq.D}
\begin{align}
\frac{d \W_D}{dt} =&-k(x_D-x_1)\dot x_1\times \Pi (t), \\
K_D(t) =&\frac{1}{2} m \dot x_D \times \Pi (t), \\
U_D (t)=&\frac{1}{2} k(x_D-x_1)^2 \times \Pi (t),
\end{align}
\end{subequations}
with $\Pi(t)$ the probability that the cantilever remains in its initial well until time $t$ given by: 
\begin{align}
\Pi(t)&= e^{-\int_0^t \Gamma(u) du}, \label{Nt}
\end{align}
where $\Gamma$ is the escape rate expressed in Eq.~\ref{EqGamma}.

\section{Kinetic temperature} \label{sec:KinT}

We define the kinetic temperature $T$ of the first deflection mode of the system through the velocity variance $\sigma_v^2=\langle v^2 \rangle-\langle v \rangle^2$:
\begin{align}
T=\frac{m}{k_B} \sigma_v^2 .\label{kinetic_temp}
\end{align}
The above definition can be reframed using the average kinetic energy $\langle K \rangle =\frac{1}{2} m \langle v^2 \rangle $, after introducing the deterministic kinetic energy contribution, $K_D$:
\begin{align}
T=\frac{2m}{k_B} (\langle K \rangle-K_D).
\end{align}

\section{SE and RE models}

We present in this section the main steps of the SE model detailed and demonstrated in Ref.~\onlinecite{Dago-2022-PRL}. 
The first step of the SE model consists in obtaining a differential equation for the kinetic temperature evolution $T(t)$ during the first step of a single erasure in a potential driving $U_1(x,x_1(t))= \frac{1}{2}k(| x | -x_1(t))^2$. This differential equation is given by the time derivative of the energy balance equation \ref{EqBalance}: 
\begin{align}
\frac{d\langle  E  \rangle}{dt} &= \frac{\partial \langle E  \rangle}{\partial T}\dot T + \frac{\partial \langle E \rangle}{\partial x_1}\dot x_1=\frac{d\langle \Q \rangle}{dt}+\frac{d\langle \W \rangle}{dt}.\label{eq.EqE}
\end{align}
The second step consists in giving the expressions of all the terms involved in the above equation. Introducing $\V=1+\erf\left(\sqrt{\frac{k}{2k_BT}} x_1\right)$, we can write: 
\begin{subequations} \label{eq.E}
\begin{align}
\frac{d\langle \Q \rangle}{dt}=& \frac{\omega_0}{Q}\big(2 K_D  + k_B T - k_B T_0\big) \label{eq.QE} \\
\frac{d\langle \W \rangle}{dt} =& \frac{d \W_D}{dt} -  k_B T \frac{\partial \ln \V}{\partial x_1} \dot{x}_1\label{eq.W}\\
\langle E \rangle =& K_D + U_D +k_B T+ k_B T^2 \frac{\partial \ln \V}{\partial T}
\end{align}
\end{subequations}
where $\W_D$, $K_D$ and $U_D$ are respectively the deterministic work, kinetic and potential energy given in Eqs.~\ref{eq.D}. 
Combining Eq.~\ref{eq.EqE} and Eqs.~\ref{eq.E}, and knowing the driving $x_1(t)$ we obtain a first order differential equation for the temperature $T(t)$ that is numerically solvable. We can then straightforwardly deduce the work and heat. Let us also point out that this model also describes the relaxation process after the end of the driving [$x_1(t)=0$].

The RE model consists in repeating this procedure for each erasure but starting with updated initial conditions each time. Let us assume for example that the temperature after the i$^\mathrm{th}$ erasure is $\tilde{T}_i>T_0$. Then the RE model will compute the temperature after the (i+1)$^\mathrm{th}$ erasure by applying the SE model with $T(0)=\tilde{T}_i$ when solving the differential equation resulting from Eq.~\ref{eq.EqE} and Eqs.~\ref{eq.E}.  

\newpage

\bibliography{repeateduse}

\end{document}